\documentclass[aps,prl,twocolumn]{revtex4}
\usepackage[dvipdfm]{graphicx} 
\usepackage{bm}
\usepackage{booktabs}
\usepackage{amsmath,amsthm,amssymb}
\usepackage{ascmac}
\newlength{\minitwocolumn}
\begin{document}

\title{Giant magnetoresistance in the junction of two ferromagnets on the surface of diffusive topological insulators}
\author{Katsuhisa Taguchi$^1$, Takehito Yokoyama$^2$, and Yukio Tanaka$^1$ }

\affiliation{$^1$Department of Applied Physics, Nagoya University, Nagoya, 464-8603, Japan \\
$^2$Department of Physics, Tokyo Institute of Technology, Tokyo, 152-8551, Japan}
\date{\today}
%
\begin {abstract}
We reveal the giant magnetoresistance induced by the spin-polarized current in the ferromagnet (F$_1$)/topological insulator (TI)/ferromagnet (F$_2$) junction, where two ferromagnets are deposited on the diffusive surface of the TI. 
We can increase and reduce the value of the giant magnetoresistance by tuning the spin-polarized current, which is controlled by the magnetization configurations.
The property is intuitively understood by the non-equilibrium spin-polarized current, which plays the role of an effective electrochemical potential on the surface of the TI.
\end{abstract}

\maketitle

Spintronics aims to control the charge transport by using spin degrees of freedom\cite{rf:Zutic04,rf:Baibich88,rf:Julliere75,rf:Maekawa82,rf:Moodera95}, and also to manipulate the spin-polarization by using the charge current\cite{rf:Datta90,rf:Nitta97,rf:Dfyakonov71,rf:Murakami03,rf:Kato04,rf:Wunderlich05}.  
The former is applicable for the giant magnetoresistance (GMR) and tunnel magnetoresistance (TMR) in the junction of a metal or an insulator sandwiched by two ferromagnets. The resistance depends on the direction of the magnetizations. 
The latter, the control of the spin-polarization, has been discussed in metals, semiconductors, in the presence of the spin-orbit interaction, which can easily induce spin polarization by the electric field.
Spin-orbit interaction, which plays a dominant role in the manipulation of the spin polarization, can also play an important role in the control of the charge transport.

Topological insulators (TIs) are candidate materials where the spin-orbit interaction enhances the magnitude of both charge and spin currents, where  
we can also expect anomalous electromagnetic phenomena stemmed from topological features of bulk electronic properties\cite{rf:Hasan10,rf:Qi,rf:Ando13}.  
Nowadays, to clarify the electromagnetic phenomena specific to TIs is believed to serve as a guide to fabricate future magneto-electronic devices.
TIs are gapped in bulk and have gapless surface state, which is dubbed as the helical surface state, in which the spin configuration and momentum are locked by spin-orbit interaction\cite{rf:Hasan10,rf:Qi,rf:Ando13}.

In the helical surface state, persistent pure spin-polarized current without dissipation is generated.
This spin-polarized current could be useful for GMR and TMR, which lead to novel magnetic devices. 
Preexisting works\cite{rf:Yokoyama10R,rf:Mondal10,rf:Kong11,rf:Ma12} have 
predicted anomalous charge transport features by the spin-polarized current in the junction of ferromagnets on the surface of 
TIs in the ballistic transport regime. 
However, the actual charge transport on the surface of TIs may be in the diffusive regime with impurity scattering. 
Thus, a study of charge transport due to the helical surfaces state is necessary in a realistic situation.

In this letter, we study charge transport between two-dimensional ferromagnet (F$_1$)/topological insulator (TI)/ferromagnet (F$_{2}$) junction, where two ferromagnets are deposited on the surface of TI. 
We find anomalous properties of GMR, which have never appeared in conventional metallic ferromagnet junctions. 
The GMR includes not only the Ohmic resistance term but also anomalous one (defined in Eq.(16)) depending on two magnetizations.
The later term is proportional to the spin polarization, which stems from the injected and extracted spin-polarized current.  
By tuning the configuration of magnetizations in the F$_{1}$ and F$_{2}$, these two terms cancel each other and gigantic reduction of magnetoresistance becomes possible. 
The present feature may serve as a guide to fabricate future magneto-electronic devices based on TIs.

Hereafter, we will consider a ferromagnet F$_{1}$/TI/ferromagnet F$_{2}$ junction on the surface of TI in the presence of the applied charge current (Fig.1 (a)).  We assume that the directions of magnetizations of F$_{1}$ and F$_{2}$ are along the $y$-axis and the direction of the charge current is along the $x$-axis. 
Then, the charge current induces the spin polarization $s^y$ by the spin-orbit interaction on the surface of TI, and $s^y$ obeys the spin-diffusion equation as\cite{rf:Burkov10,rf:Schwab11}
\begin{align} \label{eq:1}
\dot{s}^y & = \frac{3D}{2}\nabla^2_x s^y -  \frac{s^y}{\tau} + \frac{v_F}{2e} \nabla_x N_e,
\end{align}
where $N_e$, $D \equiv v_F \ell/ 2$, $v_F$, $\ell$, and $\tau$ are charge density, diffusion constant, Fermi velocity, mean free path, and 
transport relaxation time, respectively.  
The charge current drives the spatial gradient of the charge density $\nabla_x N_e$ as 
\begin{align} \label{eq:2}
\bm{j} &= - D \bm{\nabla} N_e + e v_F (\hat{\bm{z}}  \times \bm{s}). 
\end{align}
In the following discussion, we simply assume that the applied current does not have spatial dependence and is independent of $x$ ($\bm{j} =I \hat{\bm{x}} $).
The charge current, in addition, generates the spin polarization on the surface of TI by the spin-polarized current.
 The spin polarization is determined by the boundary conditions at the interface of the left-side ($x=-L/2$) between F$_1$ and TI and the right side ($x=L/2$) between TI and F$_2$, where $L$ is the distance between the two electrodes.
We use the boundary conditions at both interfaces as follows:
The spin-polarized current $\frac{\eta_1 I}{e}$ flows from F$_1$ into TI at the left interface $x=-L/2$, where $\eta_1$ is the the degree of spin polarization of the magnetization of F$_1$\cite{rf:Burkov10}.  
$\eta_{1}=1 (-1) $ is defined such that the direction of magnetization is positive (negative) along the $y$-axis and the current is fully spin-polarized. 
At the right interface, the spin-polarized current $\frac{\eta_2 I}{e}$ is extracted from TI into F$_{2}$. 
$\eta_2$ represents the direction of the magnetization F$_2$ and $\eta_2$ is defined in the same way.
On the other hand, the spin current $j_{{\rm{s}},x}^y$ on the surface of TI is given by 
\begin{align}
j_{{\rm{s}},x}^y= - \frac{3}{2}D \nabla_x s^y  -\frac{1}{2e}v_F N_e.
\end{align} 
The first and the second terms represent the non-equilibrium and equilibrium component of spin-polarized currents, respectively. 
The equilibrium component does not appear explicitly in the boundary condition\cite{rf:B.C,rf:Takahashi08}. 
Thus, the boundary conditions are given by \cite{rf:Burkov10}  
\begin{align}
 \label{eq:BDL}
 - \frac{3}{2}D \nabla_x s^y \bigr|_{x=-\frac{L}{2}} &= \frac{\eta_1 I}{e}, \\ 
 \label{eq:BDR}
 - \frac{3}{2}D \nabla_x s^y \bigr|_{x=\frac{L}{2}} &= \frac{\eta_2 I }{e}.
\end{align} 
\par

The spin polarization is obtained from Eqs. (\ref{eq:1})-(\ref{eq:BDR}) as
\begin{align} \label{eq:6}
s^y  = s^y_{1} + s^y_{2}, 
\end{align}
with
\begin{align} \label{eq:6-1}
   s^y_1 = &   \frac{I}{2 ev_F}  \frac{\kappa \ell \left\{ \eta_1 \cosh{ [ \kappa (\frac{L}{2}- x) ]}  -\eta_2 \cosh{ [ \kappa (\frac{L}{2}+ x) ]}  \right\}} {\sinh(\kappa L) }, 
     \\ \label{eq:6-2}
   s^y_2   = & - \frac{I}{2ev_F}, 
\end{align}
and $\kappa \equiv \frac{2}{\ell}\sqrt{\frac{2}{3}} $.
$s^y_1$ depends on the position $x$ ($-L/2\leq x \leq L/2$), $\eta_1$ and $\eta_2$, and it decays exponentially  with the diffusion length scale of 
$\kappa^{-1}$ from the boundaries.
$s^y_2$ is only proportional to $I$ independent of $x$. 
When $\eta_2=0$, the spin polarization is consistent with that in Ref. 19\cite{rf:spin}.
The physical meaning of $s^y_1$ is the spin polarization due to the spin-polarized current, which is given by $- \frac{3}{2}D \nabla_x s^y$ [Eqs. (\ref{eq:BDL})-(\ref{eq:BDR})]. 
The expression of $s^y_2$ means that the spin polarization generated by the spin-orbit interaction on the surface of TI\cite{rf:Raghu10}, which is independent of the magnetizations.  
\begin{figure}\centering
\includegraphics[scale=.8]{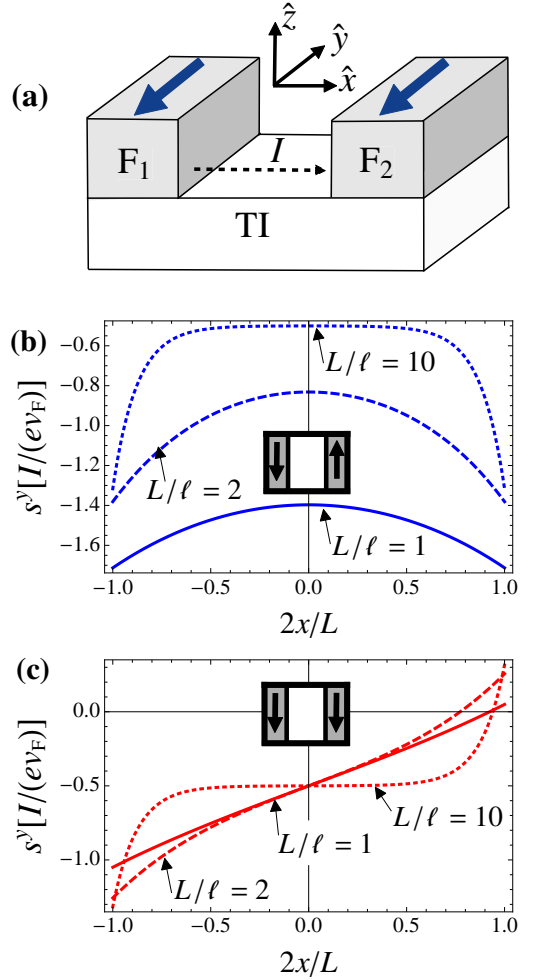}
\caption{
(Color online) (a) A schematic illustration of the F$_1$/TI/F$_2$ junction, where the arrow from F$_1$ to F$_2$ shows the applied charge current and the blue arrows represent the direction of the magnetization in F$_1$ and F$_2$.
(b)-(c) The spin polarization $s^y$ for various magnetization configurations with $L/\ell =1$ (solid line), $2$ (dashed line), and $10$ (dotted line), where $L$ is the length between two ferromagnets and $\ell$ is the mean free path on the surface of TI. 
The direction of the magnetization F$_2$ is antiparallel (b) or parallel (c) to that of F$_1$ (inset (b) or (c)).
}\label{fig:fig1}
\end{figure}

The spin polarization $s^{y}$ is plotted in Fig. 1 (b)-(c).
First, we discuss the case where magnetizations of F$_{1}$ and F$_{2}$ are antiparallel to each other with $(\eta_1, \eta_2)=(-1,1)$. 
In the antiparallel magnetization configuration, the magnitudes of $s_{y}$ decay from the interface at $x=\pm L/2$, and $s_{y}$ has a nonzero value for all $x$. 
$s_{y}$ has a strong spatial dependence for $-L/2<x<L/2$ with $L/\ell \sim 1$ and $L/\ell \sim 2$, while it becomes almost constant except for the vicinity of $x=\pm L/2$ with $L/\ell \sim 10$.  
$s^{y}$ is an even function of $x$ for all $L$ in the antiparallel magnetization configuration (see Fig. 1 (b)), while, in the parallel case [$(\eta_1, \eta_2)=(-1,-1)$], $s^{y}$ is not an even function of $x$ (see Figs. 1 (c)), because $s_1^y$ and $s_2^y (=$const$)$ are an odd and even function of $x$, respectively. 
\par

$s^{y}$ contributes to the voltage drop $V$ between F$_{1}$ and F$_{2}$ on the surface of TI due to the spin-orbit interaction\cite{rf:Burkov10}. 
$V$ is defined by \cite{rf:voltage} 
\begin{align}
V & =  -\frac{1}{e^2 \nu} \int_{-L/2}^{L/2} \frac{dN_e}{dx} dx,
\end{align}
where $\nu = \frac{\mu }{2\pi v_F^2}$ is the density of states on the surface state of TI at chemical potential $\mu = v_F k_F$. 
From Eq. (\ref{eq:2}), the voltage drop can be expressed by 
\begin{align} \label{eq:voltage-drop1}
V & =  V_{\rm{O}} +V_{\rm{s}}, 
\end{align}
with 
\begin{align} \label{eq:voltage-drop1}
V_{\rm{O}} & =  \frac{2}{e^2 \ell \nu v_F } \int_{-L/2}^{L/2}  dx ( I +e v_F s^y_2), \\ \label{eq:voltage-drop2}
V_{\rm{s}} & =  \frac{2}{e \ell \nu } \int_{-L/2}^{L/2}  dx s^{y}_1. 
\end{align}
In the above, the voltage drop $V_{\rm{O}}$, which is given by the charge current and constant spin polarization due to $I$, obeys the Ohm's law.
The spin polarization $s^y_2$ in the second term of $V_{\rm{O}}$ is negative in the presence of the charge current flowing towards positive $x$-axis, and hence $s^y_2$ has an opposite contribution to the first term of $V_{\rm{O}}$ in Eq. (\ref{eq:voltage-drop1}). 
On the other hand, the voltage drop $V_{\rm{s}}$ in Eq. (\ref{eq:voltage-drop2}) is given by the spin polarization $s^y_1$, which depends on the magnetization configuration [see Eq. (\ref{eq:6-1})].
Since $V_{\rm{s}}$ is determined by the magnetization configurations, the resulting  $V_{\rm{s}}$ only depends on the boundary condition of spin-polarized current and no more obeys the Ohm's law. 
$V_{\rm{s}}$ can be regarded as a boundary term and is obtained as $V_{\rm{s}}   =  \frac{ \eta_1 -\eta_2}{e^2 \nu v_F} I $.
Finally, the total voltage drop $V$ is given by 
\begin{align}\label{eq:total-voltage}
V & =  \frac{1}{e^2 \nu v_F}  \left( \frac{L}{\ell} + \eta_1 -\eta_2 \right) I. 
\end{align}
The first term in Eq. (\ref{eq:total-voltage}) obeys the Ohm's law and is proportional to the length $L$ of TI. 
The second (third) term is independent of $L$, and is proportional to $\eta_1$ ($\eta_2$).
The coefficient $\frac{1}{e^2 \nu v_F}$ is proportional to the Sharvin resistance in the two-dimension\cite{rf:Sharvin}.

\begin{figure}\centering \includegraphics[scale=.51]{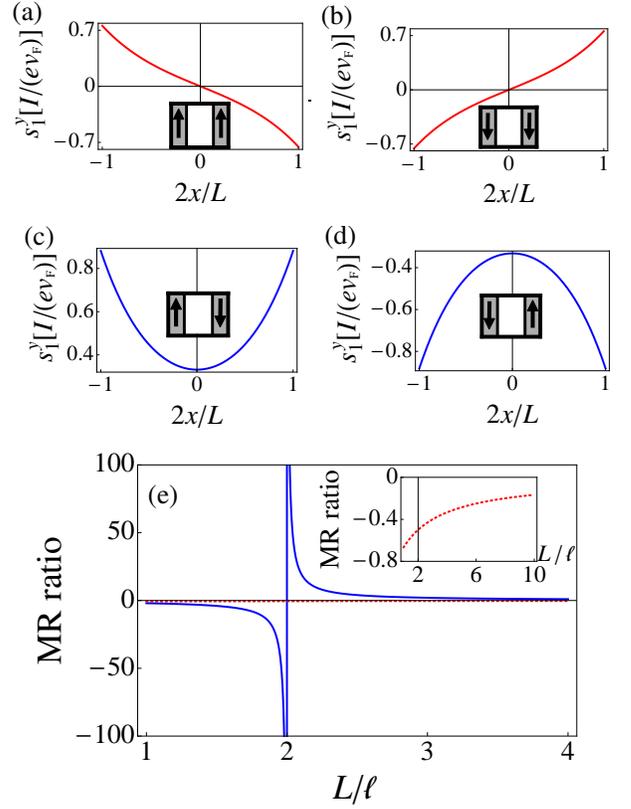}  
\caption{(Color online) The spin polarization, $s^y_1$, for antiparallel [(a) and (b)] and parallel [(c) and (d)] magnetization configurations with $L/\ell=2$. 
 The schematic magnetization configurations in F$_1$/TI/F$_2$ junctions are shown in the insets. 
(e) Magnetoresistance (MR) ratio  as a function $L$. 
The blue solid (red dashed) line represents 
MR$_{\downarrow} = \frac{R_{\downarrow \downarrow}-R_{\downarrow \uparrow}}{R_{\downarrow \uparrow}} $ 
(MR$_{\uparrow} = \frac{R_{\uparrow \uparrow }-R_{\uparrow \downarrow}}{R_{\uparrow\downarrow}}$).
The enlarged plot of MR$_{\uparrow} $ is shown by the dashed line in the inset.  
}
\label{fig:fig2}\end{figure}
The magnetoresistance $V/I = \frac{1}{e^2 \nu v_F} ( \frac{L}{\ell} +\eta_1 -\eta_2 )$ depends on the configurations of the magnetizations in F$_1$ and F$_2$: 
\begin{align}\label{eq:antiparallel}
R_{\sigma \sigma'} & =  R_{\rm{O}} + R_{\rm{s},\sigma \sigma'}, 
\\ \label{eq:parallel-n}
R_{\rm{O}} & = \frac{L}{e^2 \nu v_F \ell}, 
\\ \label{eq:parallel-p}
R_{\rm{s},\sigma \sigma'}&= \frac{1}{e^2 \nu v_F} (\eta_{1}-\eta_{2})
\end{align}
where $\sigma$ and $\sigma'$ denote spin indices with $\sigma=\uparrow (\downarrow)$ and $\sigma'=\uparrow (\downarrow)$.
$\eta_{1}=1 (-1)$ and $\eta_{2}=1(-1)$ correspond to $\sigma=\uparrow(\downarrow)$ and $\sigma'=\uparrow(\downarrow)$, respectively. 
The magnetoresistance has two terms: Ohmic resistance term ($R_{\rm{O}}$) and the boundary term ($R_{\rm{s},\sigma\sigma'}$).
$R_{\rm{O}}$ does not depend on the magnetization configurations.
$R_{\rm{s},\sigma\sigma'}$ depends on the magnetization configurations.
For the parallel configuration, the boundary term is absent.
On the other hand, for the antiparallel configurations, the boundary term exists.
Moreover, in the antiparallel configuration, the boundary term has a negative (positive) sign when the direction of the magnetization F$_1$ and F$_2$ are negative (positive) and positive (negative) along the $y$-axis, respectively.
Whether $R_{\rm{s}, \sigma\sigma'}$ is zero or nonzero can be seen from the difference of the spatial dependences of $s^y_1$ for $L/\ell =2$ in Fig. 2 (a)-(d).
$s^y_1$ is even (odd) function of $x$ in antiparallel (parallel) configuration (see inset in Fig. 2 (a)-(d)).
Since $R_{\rm{s}, \sigma\sigma'}$ is obtained by $V_{\rm{s}}/I \propto \int^{L/2}_{-L/2} dx s_1^y$, it is nonzero (zero) in the antiparallel (parallel) configurations of the magnetizations.
The sign of the $R_{\rm{s}, \sigma\sigma'}$ term can be seen from the sign of $s^y_1$ in Fig. 2 (c) and (d), and is positive (negative) in $R_{\uparrow \downarrow}$ ($R_{\downarrow \uparrow}$).

These properties of the magnetoresistance are quite different from those of the conventional magnetoresistance\cite{rf:Zutic04,rf:Baibich88,rf:Julliere75,rf:Maekawa82,rf:Moodera95}:
The magnetoresistance is only proportional to the relative angle between the directions of the two magnetizations (i.e., $R_{\uparrow\uparrow} = R_{\downarrow\downarrow}$ and  $R_{\uparrow\downarrow} = R_{\downarrow\uparrow}$).
The magnitude of the magnetoresistance is usually given by $R_{\uparrow\uparrow} < R_{\uparrow\downarrow}$.
On the other hand, the magnetoresistance in Eqs. (\ref{eq:antiparallel})-(\ref{eq:parallel-p}) behaves in a drastically different way. 
There are two conditions to determine the magnitudes of magnetoresistance $R_{\sigma\sigma'}$: 
Whether the magnetization configuration is  parallel or antiparallel, and whether the direction of the magnetizations F$_1$ and F$_2$ are positive and negative or negative and positive along the $y$-axis in the antiparallel configuration, respectively.
$R_{\sigma\sigma'}$ satisfies $R_{\uparrow\uparrow} = R_{\downarrow\downarrow}$ and $R_{\downarrow\uparrow} < R_{\uparrow\downarrow}$, and we can obtain $R_{\downarrow\uparrow} < R_{\uparrow\uparrow} < R_{\uparrow\downarrow}$ from Eqs. (\ref{eq:antiparallel})-(\ref{eq:parallel-p}).
Thus, the relation $R_{\downarrow\uparrow} < R_{\uparrow\uparrow}$ is different from that of the conventional case, although $R_{\uparrow\uparrow} = R_{\downarrow\downarrow}$ and  $R_{\uparrow\uparrow} < R_{\uparrow\downarrow}$ are the same properties as that of the conventional magnetoresistance.
Moreover, $R_{\downarrow\uparrow}$ is lower than the Ohmic resistance, and becomes zero for  $L/\ell =2$\cite{rf:Burkov10}.
The anomalous property of the magnetoresistance, $R_{\downarrow\uparrow}$, could be applicable for an charge transport with low Joule heating.
\par

We will consider the magnetoresistance ratio, MR$_{\downarrow}$ (MR$_{\uparrow}$), when the direction of the magnetization F$_2$ is free and F$_1$ is fixed along the negative (positive) $y$-axis.
These magnetoresistance ratios are given by 
\begin{align}\label{eq:MR-1} 
\text{MR}_{\downarrow} & \equiv \frac{R_{\downarrow \downarrow}-R_{\downarrow \uparrow}}{R_{\downarrow \uparrow}} = \frac{2}{L/\ell - 2}, \\ \label{eq:MR-2}
\text{MR}_{\uparrow} & \equiv \frac{R_{\uparrow \uparrow }-R_{\uparrow \downarrow}}{R_{\uparrow\downarrow}} = \frac{-2}{L/\ell + 2}. 
\end{align}
$\text{MR}_{\downarrow}$ and $ \text{MR}_{\uparrow} $ depend on $L/\ell$.
The value of $\text{MR}_{\downarrow}$ is negative for $L<2\ell$ and is positive for  $L>2\ell$, and diverges at $L/\ell =2$ (see Fig. 2(b)) since $R_{\downarrow \uparrow}$ is zero.
The value of $\text{MR}_{\uparrow}$ is negative for all $L/\ell$ (see the inset of Fig. 2(b)).
The magnitude of $\text{MR}_{\uparrow}$ and $\text{MR}_{\downarrow}$ are  49$\%$ and 1000$\%$ in $L/\ell =2.1$, respectively.
The gigantic magnetoresistance ratio $\text{MR}_{\downarrow}$ stems from the boundary term $R_{{\rm{s}}, \sigma\sigma'}$ in the surface state of the TI.

Finally, we show that the value of $V_{\rm{s}}$ can be intuitively understood from the boundary conditions of the spin-polarized current.
$s_1^y$ of $V_{\rm{s}}$ can be rewritten by Eqs. (\ref{eq:1}) and (\ref{eq:2}) as\cite{rf:derivation} 
\begin{align}
s_1^y & = \frac{3}{4}\tau D \nabla_x^2 s_1^y.
\end{align}
By using the above equation, $V_{\rm{s}}$ in Eq. (\ref{eq:voltage-drop2}) is given by the integration of the spatial gradient of the non-equilibrium spin-polarized current $-\frac{3}{2}D \nabla_x s^y$, and the value of $V_{\rm{s}}$ is proportional to the difference of the non-equilibrium spin-polarized current at the interfaces:
\begin{align}\label{eq:difference of spin}
V_{\rm{s}} & = \frac{-1}{e\nu v_F} \int_{-L/2}^{L/2} \nabla_x \left(-\frac{3}{2} D \nabla_x s_1^y \right) dx \\ \label{eq:difference of spin2}
	& = \frac{-1}{e\nu v_F}  \biggl(-\frac{3}{2} D \nabla_x s_1^y\bigr|_{x\to-\frac{L}{2}}^{x\to\frac{L}{2}} \biggr)
	= \frac{\eta_1 -\eta_2}{e^2 \nu v_F} I.
\end{align}
In the electromagnetism, the voltage drop is generated by electrochemical potential $\phi$ as $\int dx \nabla_x \phi$ and the electric field is given by $-\nabla_x \phi$.
Therefore, we can regard the non-equilibrium spin-polarized current as an electrochemical potential and its gradient as an effective electric field, which drives the anomalous voltage drop in the diffusive surface of TIs.

In conclusion, we studied the anomalous magnetoresistance in the F$_1$/TI/F$_2$ junction by using the spin-diffusion equation on the surface of TIs.
It is found that the magnetoresistance includes the boundary term, which is absent in the usual ferromagnetic junction.
The boundary resistance is triggered by the spin polarization of the current, which is induced by both the spin-polarized current injected from F$_1$ into the TI and extracted from the TI into F$_2$. 
The boundary resistance plays the role of the reduction (enhancement) to the Ohmic term when the directions of the two magnetizations are negative (positive) and positive (negative) along the $y$-axis.
Recently, a fine quality film of the TI comes to be able to fabricate\cite{rf:Taskin12a,rf:Taskin12b}. Also, it has been demonstrated that the ferromagnetism at ambient temperature can be induced in the Mn-doped Bi$_2$Te$_3$ by the magnetic proximity effect through deposited Fe overlayer\cite{rf:Vobornik}.
By using the film of the TI, the boundary resistance can be experimentally demonstrated and would be applicable for a charge transport with low energy consumption in future.

This work was supported by a Grant-in-Aid for Young Scientists (B) (No. 22740222, No. 23740236)) and by a Grant-in-Aid for Scientific Research on Innovative Areas ''Topological Quantum Phenomena'' (No. 22103005, No. 25103709) from the Ministry of Education, Culture, Sports, Science and Technology, Japan (MEXT).
K.T. acknowledges the support by the JSPS.


\begin{references}
\bibitem{rf:spin} $s_1^y$ can be also expressed as $s_1^y =      \frac{\sqrt{\frac{2}{3}}\frac{I}{ev_F}} {\sinh(2\sqrt{\frac{2}{3}} \frac{L}{\ell})}   [\eta_1 \cosh{ ( \sqrt{\frac{2}{3}}\frac{L- 2x}{\ell}  )} - \eta_2  \cosh{ ( \sqrt{\frac{2}{3}}\frac{L+2x}{\ell} )} ] $.
\bibitem{rf:voltage} Derivation of the voltage drop: 
By using the charge density $N_e (\mu ) = e \sum_k f(\epsilon_k -\mu)$, the difference $\delta N \equiv N_e (\mu + eV) -N_e (\mu) $ at low temperature is given by 
$\delta N_e  = e \sum_k [ f(\epsilon_k -\mu -eV) -f(\epsilon_k -\mu)] =  e^2 \nu V $.
$\delta N$ is also written, when there is the voltage drop $V$ between $x=-\frac{L}{2}$ and $x=\frac{L}{2}$, as  
$\delta N_e =  N_e (x=-\frac{L}{2}: \mu + eV)  -N_e (x=\frac{L}{2}: \mu) = - \int^{\frac{L}{2}}_{-\frac{L}{2}} dx' \frac{d N_e (x)}{dx'}$.
\bibitem{rf:derivation}The spin-diffusion equation in Eq. (1) is rewritten by  Eq. (2) as $\frac{3D}{2} \nabla^2_x s^y - \frac{2}{\tau}s^y - \frac{I}{e\ell} =0$. To substitute $s^y = s'^y - \frac{I\tau}{2e\ell}$ into the above equation, we obtain $\frac{3D}{2} \nabla^2_x s'^y = \frac{2}{\tau}s'^y$. Now $s'^y$ is equal to $s_1^y$, because of $-\frac{I\tau}{2e\ell} = s_2^y$ and $s'^y = s^y + \frac{I\tau}{2e\ell} = s^y - s_2^y$ from Eq. (6).
\end{references}
\end{document}